# Thermal boundary conductance and thermal conductivity strongly depend on nearby environment

Khalid Zobaid Adnan, Tianli Feng*

Department of Mechanical Engineering, University of Utah, Salt Lake City, Utah 84112, USA

**Corresponding Authors**

*Email: tianli.feng@utah.edu

**Abstract:**
At the nanoscale, the thermal boundary conductance (TBC) and thermal conductivity are not intrinsic properties of interfaces or materials but depend on the nearby environment. However, most studies focused on single interfaces or superlattices, and the thermal transport across heterostructures formed by multiple different materials is still mysterious. In this study, we demonstrate how much the TBC of an interface is affected by the existence of a second interface, as well as how much the thermal conductivity of a material is affected by the nearby materials. Using Si and Ge modeled by classical molecular dynamics simulations, the following phenomena are demonstrated. (1) The existence of a nearby interface can significantly change the TBC of the original interface. For example, by adding an interface after Si/Ge, the TBC can be increased from 400 to 700 MW/m²K. This is because the nearby interface serves as a filter of phonon modes, which selectively allows particular modes to pass through and affect the TBC of the original interfaces. This impact will disappear at the diffusive limit when the distance between interfaces is much longer than the phonon mean free path so that phonon modes recover equilibrium statistics before arriving at the second interface. (2) The thermal conductivity of a material can be significantly changed by the existence of neighboring materials. For example, a standalone 30-nm-thick Si's thermal conductivity can be increased from 50 to 280 W/mK, a more than 4-fold increase, beating the bulk thermal conductivity of Si, after being sandwiched between two Ge slabs. This is because the Ge slabs on the two sides serve as filters that only allow low-frequency phonons to transport heat in Si, which carry more heat than optical phonons. This work opens a new area of successive interface thermal transport and is expected to be important for nanoscale thermal characterization and thermal management of semiconductor devices.

# I. INTRODUCTION

Semiconductor heterostructures are core components in a wide range of applications including microprocessors[1–3] in computing devices, converters[4,5] and inverters[6] in power electronics, radio-frequency[7,8] components in communication systems, light emitting diodes (LEDs)[9,10] and laser diodes[11,12] in opto-electronics, sensors[13–15] and controllers[16] in automotive technologies, and critical elements in industrial machinery and medical imaging systems[17–19]. In most applications, the heterostructures are made of multiple layers of different semiconductor materials, and the heterostructure interfaces account for a substantial part of the near-junction thermal resistance. Understanding and managing heat transfer at these material interfaces is pivotal for enhancing the performance and reliability of these systems [20–24].

Interfacial thermal resistances of many structures have been measured experimentally[25–29] and simulated theoretically[30,31]. However, most studies have focused on heat transport across a single interface or in superlattices and random multilayers with numerous interfaces. A few studies noticed the coupling between two 1D/2D junctions in simulations[32] but did not find it in double metal/nonmetal interfaces in experiment[33]. Although it is well-known that interfaces can affect each other, it remains unclear under which conditions and for what types of interfaces these effects occur, how strong they are, and how to predict the coupling - even in the simplest cases, such as when only two interfaces are present.

The heat transport across two successive interfaces differs from flowing through two independent interfaces because the interfaces can affect each other. This interaction has two primary effects: (1) The first interface acts as a phonon filter, selectively allowing specific phonon modes to pass through while blocking others. Consequently, only the phonon modes permitted by the first interface can reach the second interface. As a result, the second interface's TBC depends on the first interface's characteristics and vice versa. (2) The phonons reflected by the second interface can travel back to the first interface and influence its TBC. The sketch in Fig. 1 shows an example of the interplay between nearby interfaces.

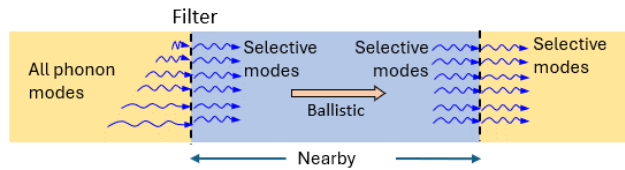

Fig. 1. Impact of nearby environment on thermal transport. Each interface serves as a phonon mode filter, which only allows selective modes to pass through. The phonon modes that can reach an interface are dependent on whether they can pass another interface. The figure only shows the impact of one on the other, but it should be noted that the impact is mutual.

The inter-dependent thermal transport across two interfaces occurs when the distance between the two interfaces is shorter than the phonon mean free path (MFP). In such cases, phonons can travel ballistically between the two interfaces within the film without experiencing significant scattering or dissipation. This case is commonly seen in practical applications. For example, depending on the design of semiconductor devices, the thickness of the semiconductor layers in a transistor can range from nanometers to microns[34–46]. Despite the ubiquity and significance of successive interfacial thermal transport, research in this area has been rare. Past and current research efforts focus on studying the thermal conductance of single interfaces[20,47–50], assuming that interfacial thermal transport is independent (except for superlattices). As a result, many physical mechanisms of successive interfacial thermal transport are unclear.

It is worth noting that interdependent thermal transport across two successive interfaces discussed here differs from the coherent and incoherent phonon thermal transport studied for superlattices or random multilayers in the literature[51–54]. First, the superlattices or random multilayers require only two materials to form many but the same interfaces, being spaced equally (for superlattices) or non-equally (for random multilayers). In contrast, successive interfacial thermal transport studied in this work only requires two or more interfaces and does not require the same interface or a periodic thickness. Second, the superlattice and random multilayer studies focus on coherent and incoherent phonons, relying on the wave nature of phonons and their interference at multiple interfaces. The periodic lattice changes the phonon band structure, folds the Brillouin zone, and creates band gaps[55–57]. In contrast, the phonon band is not altered in successive interfacial thermal transport, and wave interference is not present or essential. Third, since the dominant phonon MFP can be very long (~μm)[58–60], but the dominant phonon wavelength for thermal conductivity is short (<10 nm)[61], the MFP impact should dominate over wavelength effect when the thickness of a layer is greater than 10 nm. Last, the superlattice studies focus on the thermal conductivity of the entire structure (composite material), while this paper focuses on the impact of the subsequent interfaces on the thermal transport of the other interfaces and regions.

In this paper, we test our hypothesis and demonstrate the interplay between two interfaces using model systems, i.e., Si/Ge vs. Si/Ge/M and Ge/Si vs Ge/Si/M. Throughout our studies, we are consistent on the type of first interface, Si/Ge or Ge/Si, and focus on the impact of the change of the second interface, Ge/M or Si/M. The remainder of this paper is organized as follows. In Section II, the nonequilibrium molecular dynamics (NEMD) simulation setups for Si, Ge, Si/Ge, Ge/Si, Si/Ge/M, and Ge/Si/M are described. The apparent and conventional thermal conductivities obtained from NEMD are defined. In section III A, the thickness-dependent and bulk thermal conductivities of standalone Si and Ge are obtained from NEMD. In Sec. III B, we study the impact of the second interface on the first interface's TBC by comparing Si/Ge and Si/Ge/M (as well as Ge/Si and Ge/Si/M) systems with varying M. Then, we select Si/Ge and Si/Ge/Si (as well as Ge/Si and Ge/Si/Ge) to study the impact of the mid-layer thickness, which is the distance between two interfaces, on the first interface's TBC. In Sec. III C, the impact of the second interface on the apparent and conventional thermal conductivity values of the first material is studied. In Sec. III D, the impact of the second interface on the conventional thermal conductivity for middle material is studied. In Sec. III E, the impact of the second interface on the total thermal resistances is discussed. In Sec. IV, the conclusions are drawn.

## II. METHODOLOGY

Figure 2 shows three representative systems, including a standalone material, a single-interface structure made of two materials, and a double-interface structure made by three materials. For all three systems, we assume the cross-section is infinitely large, and the system sizes are only limited along the heat transport direction, defined as the $x$ direction.

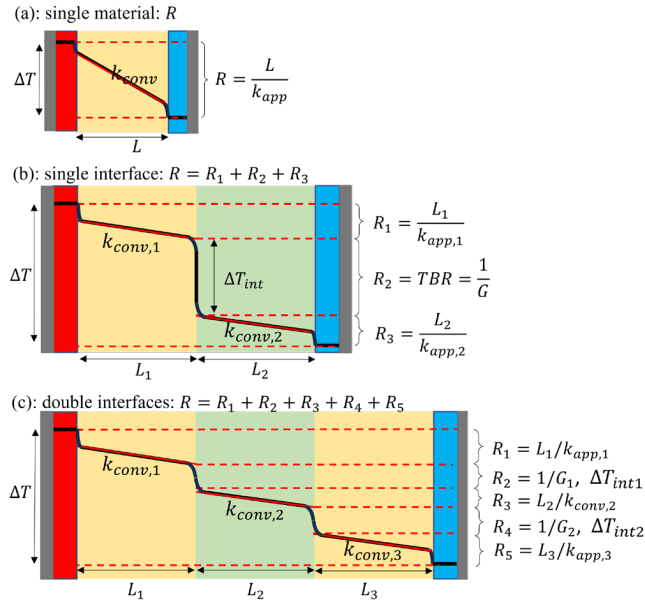

**FIG. 2.** Schematics for a material that is (a) standalone, (b) attached with a second material, and (c) attached with a third material. (a-c) are named standalone material, single-interface system, and double-interface system, respectively. The left and right grey edges for each system represent adiabatic boundary conditions. The red and blue stripes on the left and right are hot and cold thermal reservoirs, respectively.

For the standalone material in Fig. 2 (a), when applied a temperature difference $\Delta T$, a linear temperature profile will be established inside the material. If the material's thickness is not much larger than phonon MFP, temperature jumps will show up near the two ends, due to ballistic phonon transport. A conventional way to calculate the thermal conductivity[62,63] of such a thin film is to use the temperature slope in the linear region inside the material, i.e.,

$$k_{conv} \equiv \frac{q}{\nabla T_{\text{lin reg}}}, \tag{1}$$

where $q$ is the heat flow rate per unit cross-section area. $\nabla T_{\text{lin reg}}$ is the gradient of temperature in the linear region inside the material. In contrast to the conventional thermal conductivity, another way to define the thermal conductivity is the apparent thermal conductivity[64], $k_{app}$, i.e.,

$$k_{app} = \frac{q}{\Delta T/L}, \tag{2}$$

which uses the temperature difference $\Delta T$ applied to the material and the thickness of the material to calculate the effective temperature gradient. This definition is closer to experimental measurement. By this definition, the total resistance of the material is:

$$R \equiv \frac{\Delta T}{q} = \frac{L}{k_{app}}. \tag{3}$$

For the single-interface system in Fig. 2 (b), the thermal conductivity $k_{conv,1}$, $k_{conv,2}$, $k_{app,1}$, and $k_{app,2}$ are defined for materials 1 and 2, respectively, in the same ways as for standalone materials. The thermal boundary conductance is defined as

$$G_1 = \frac{q}{\Delta T_{int}}, \tag{4}$$

where $\Delta T_{int}$ is the temperature jump at the interface. With these definitions, the total resistance of the structure is the summation of the apparent resistances and thermal boundary resistance:

$$R_{tot} = \frac{L_1}{k_{app,1}} + \frac{1}{G_1} + \frac{L_2}{k_{app,2}}. \tag{5}$$

Note that the resistance of each material is calculated from its apparent thermal conductivity rather than conventional thermal conductivity, because the temperature jumps at the left and right leads need to be taken into account in the total resistance.

For the double-interface system in Fig. 2, we can define $k_{conv,1}$, $k_{app,1}$, $k_{conv,2}$, $k_{conv,3}$ and $k_{app,3}$ for materials 1, 2, and 3, respectively, in the same way as for a standalone material. But there is no $k_{app,2}$ for material 2 since the temperature jumps at the two ends of material 2 are included in the thermal boundary resistances of the two interfaces and should not be double counted. With these definitions, the total resistance of the structure is:

$$R_{tot} = \frac{L_1}{k_{app,1}} + \frac{1}{G_1} + \frac{L_2}{k_{conv,2}} + \frac{1}{G_2} + \frac{L_3}{k_{app,3}}. \tag{6}$$

The study is based on NEMD simulations using Large-scale Atomic Molecular Massively Parallel Simulator (LAMMPS)[65–67]. The lattice constant for both Si and Ge is taken as $a = 5.442$Å, which is obtained by relaxation of the structures in MD at 300 K using the Tersoff interatomic potential[68]. The cross-section lengths of all the systems are $L_y = L_z = 43.54$ Å. The timestep of NEMD simulations is set to be 1 fs. The NEMD simulation setup is similar to a prior work[68]. The left and right edge atoms, with a thickness of 1 nm, are fixed to mimic the adiabatic boundary conditions. Hot and cold reservoirs are next to the fixed edge atoms, with a reservoir thickness of 3.3 nm. Between the two reservoirs is the system. Periodic boundary conditions are imposed in the lateral directions to mimic infinite $y$ and $z$ dimensions. The structures are first relaxed under a constant volume and temperature (NVT) ensemble for 20 ns at 300 K and then shifted to a constant energy and volume (NVE) ensemble. In the NVE ensemble, the hot and cold reservoirs are kept at 320 and 280 K using the Langevin[69] thermostat and stabilize the heat current for 20 ns. The next 20 ns are recorded to extract heat flux and temperature profiles inside the systems. Langevin thermostat excites all the phonon modes equally and serves as a non-biased phonon reservoir[69–71]. The damping parameter is set to be 0.5 ps, which is appropriate to provide a stable heat current[72]. The heat sources and heat sinks for all the systems are shown in red and blue, respectively, in Fig.1. A tentative temperature profile obtained from NEMD for each system is also shown. From the temperature gradients, jumps, and heat current, we evaluate the thermal conductivity of the constituent materials and TBC at the interfaces. All the thermal conductivity and TBC reported in this paper are classical values without quantum corrections. The quantum correction will reduce the classical TBC of Si/Ge interface[73] by about 8%. The quantum correction to thermal conductivity is still under debate[74] since it is impossible to correct the temperature, phonon population, zero-point energy, phonon scattering rate, and specific heat simultaneously.

## III. RESULTS AND DISCUSSION

### A. Size-dependent thermal conductivities of Si and Ge

The size-dependent $k_{conv}$ and $k_{app}$ of standalone Si and Ge films obtained from NEMD in this work are shown in Fig. 3. Both $k_{conv}$ and $k_{app}$ increase with material thickness $L$, even at the maximum thicknesses simulated in this work, 500 and 160 nm for Si and Ge, respectively. To obtain the bulk thermal conductivity, we plot $1/k$ vs $1/L$, and find that $1/k_{conv}$ and $1/k_{app}$ converge at the same point, which corresponds to $1/k(\infty)$ or $1/k_{bulk}$. In this work, $k_{bulk}$ of Si and Ge are found to be 218±12, and 107±7 W/mK, respectively. These results are close to the literature reported bulk values, 233.4 and 93.3 W/mK, by using an approach-to-equilibrium molecular dynamics (AEMD)[75] method. The slight discrepancy in Ge thermal conductivity comes from the use of Si Tersoff potential for Ge in this work. We acknowledge that these values are higher than the experimental data[76], which are 140 and 63 W/mK, respectively. This is because (1) the Tersoff potential overestimates the phonon velocities and likely lifetimes as well, and (2) MD uses classical statistics and needs quantum corrections to compare with the experiment. This work is not focused on the absolute thermal conductivity or TBC values but on their relative changes under the impacts of neighboring interfaces. Therefore, even though the absolute value cannot match the experiment, it does not affect our self-consistent relative comparison for the exploration of physics.

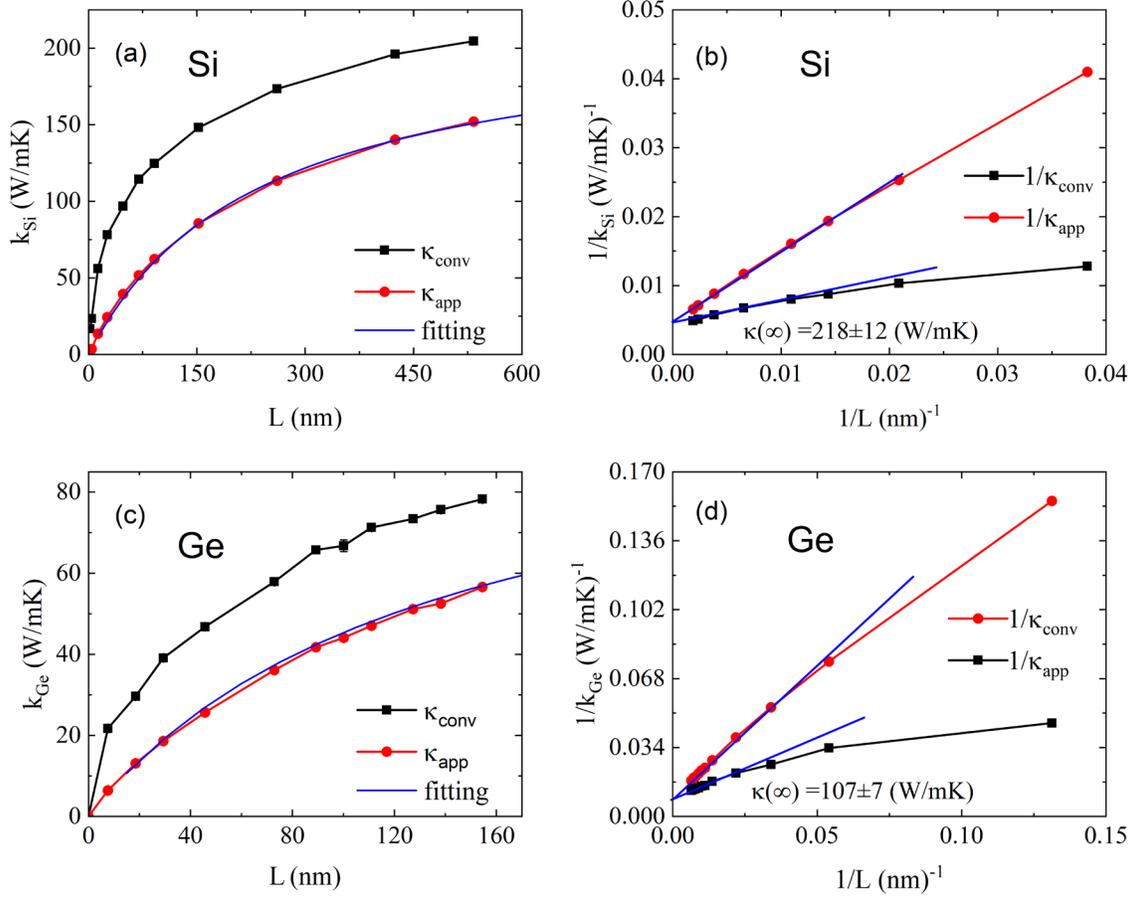

**FIG. 3**. Size effect of standalone Si and Ge films. Symbols are data obtained from NEMD simulations. The fitting curves in (a) and (c) are obtained from the grey phonon radiation model.

For a better understanding of the remaining results, it is worth estimating the phonon MFP of Si and Ge, which can be extracted from the size-dependent thermal conductivity in Fig. 3. Based on the gray phonon radiation model, the size-dependent thermal conductivity can be expressed by[64,77]

$$k_{app} = \frac{k_{bulk}}{1+\frac{4\Lambda}{3L}}. \tag{7}$$

Thus, the MFP $\Lambda$ can be obtained by linear fitting of $1/k_{app}$ vs. $1/L$ with

$$\frac{1}{k_{app}} = \frac{1}{k_{bulk}} + \frac{4\Lambda}{3k_{bulk}}\frac{1}{L}. \tag{8}$$

The obtained MFP of Si and Ge are about 178 and 102 nm, respectively. Using the obtained $\Lambda$ and $k_{bulk}$ values, we plot the calculated $k_{app}(L)$ using Eq. (7) and compare it with NEMD data in Fig. 3 (a) and (c). They agree well with each other, demonstrating the accuracy of the gray phonon radiation model. Note that these MFP values are just an estimation, which is an average of the broad-band spectrum of the real MFP. Another way to estimate the MFP is by $\Lambda = k_{bulk}/(\frac{1}{3}cv)$, which only gives ~46 nm MFP for Si [64]. This value significantly underestimates the MFP since it assumes all phonon modes have the same velocity as the sound speed $v$. We note that the MFPs estimated in this work are not used to compare with those calculated from the first principles [58,59,78] since classical Tersoff potential overestimates the thermal conductivity of Si and Ge. The MFPs here are used to understand the impacts of thickness on thermal conductivity and interfaces in the following sections.

## B. Impact of second interface on first interface's TBC

To study the impact of a second interface on the first interface's TBC, we add a third material, M, after Si/Ge, to form a Si/Ge/M double-interface system. M is an arbitrary material with the same lattice structure as Si but a different mass. In the Si/Ge/M system, there are two interfaces: the first interface, Si/Ge, and the second interface, Ge/M. The first interface's TBC is named $G_1$, throughout this paper. We study the impact of different types of second interfaces by varying M from 6 to 300 atomic unit (a.u.) mass.

Figures 4 (a-c) compare $G_1$ in Si/Ge vs. Si/Ge/M systems. We find that $G_1$ in Si/Ge/M, in most cases, is higher than the original TBC of Si/Ge. In other words, the existence of the second interface (Ge/M), in most cases, increases the first (Si/Ge) interface's TBC. In Fig. 4 (a), by fixing the lengths of Si, Ge, and M at 10 nm, and changing the mass of M only, we find that $G_1$ reaches the maximum at M=Si=28, when Si/Ge/M becomes Si/Ge/Si. At this point, $G_1$ in Si/Ge/M is 33% higher than its original value in Si/Ge. In Fig. 4 (b), we fix the lengths of Si, Ge, and M to 27 nm, and the maximum $G_1$ still occurs at M=Si=28 and is 26% higher than the single interface value. In Fig. 4 (c), when lengths are fixed at 54 nm, the maximum $G_1$ is 16% higher than the single interface value. These findings agree with our hypothesis that the interplay between interfaces is stronger when closer.

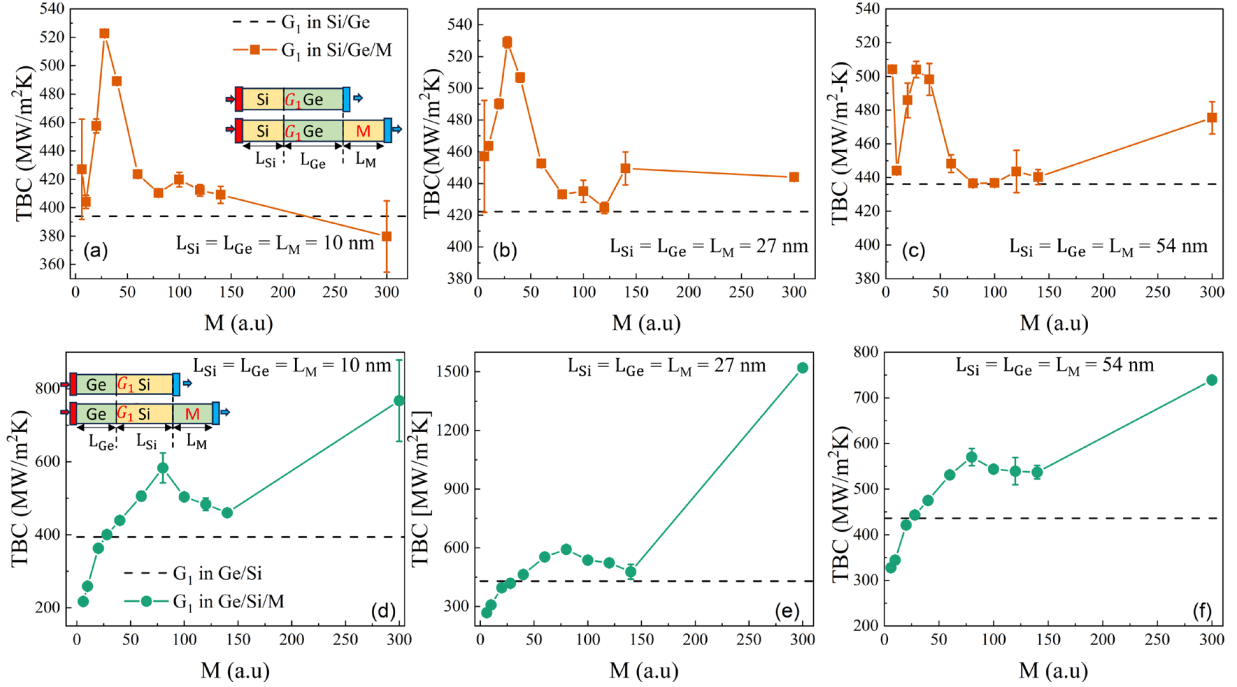

**FIG. 4**. The impact of second interface on first interface's TBC. (a-c) $G_1$ in Si/Ge and Si/Ge/M as a function of mass of M. (d-f) $G_1$ of Ge/Si and Ge/Si/M as a function of mass of M.

Figures 4 (d-f) show the Ge/Si TBC in Ge/Si vs. Ge/Si/M systems. Similar phenomena are observed: the maximum of $G_1$ occurs at M=Ge=72, when Ge/Si/M becomes Ge/Si/Ge, a symmetry system. The enhancement of $G_1$ by the existence of M is large. For example, when lengths of Si, Ge, and M are 10 nm, $G_1$ of Ge/Si/Ge is 50% higher than that of Ge/Si at the same length. It is understandable that the maximum $G_1$ occurs for the A/B/C systems when C=A. In the A/B/A system, since the A/B and B/A interfaces are identical[79,80], all the phonon modes that can pass through one interface can also pass the other. This increases the apparent phonon transmission coefficient at the other interface, and thus increases the TBC. This agrees with our hypothesis that each interface serves as a filter, and when two filters are identical, the effective transmission is boosted the most[79,80].

To explore extreme physics, we have also done simulations when M is extremely light or heavy. Since the simulations are not stable when M is too large (e.g., more than 500) or too small (e.g., less than 5), we limit M to be between 6 to 300. The results can be discussed in the following categories. (1) When M is light, e.g., M=6, $G_1$ in Si/Ge/$^6$M is higher than that of Si/Ge, but $G_1$ in Ge/Si/$^6$M is smaller than that of Ge/Si. This is understandable. In Si/Ge/$^6$M, the two side materials are both lighter than the middle material, so that each interface serves as a "good" phonon mode filter for the other, since the phonon modes that can commute through Si/Ge can likely also commute through $^6$M/Ge. In contrast, in Ge/Si/$^6$M, the two interfaces are not beneficial filters for each other. The phonon modes that can commute through Ge/Si are likely low-frequency phonons, while the modes that can commute through Si/$^6$M are likely high-frequency phonons. (2) When M is heavy, e.g., M=300, $G_1$ in Ge/Si/$^{300}$M is higher than that of Ge/Si, but $G_1$ in Si/Ge/$^{300}$M may not be higher than that of Si/Ge. The physical reason is similar to Case (1). In summary, if the third material follows the first material's phonon characteristics, e.g., the two side materials are both lighter or heavier than the middle material, then the existence of the third material can enhance the first interface's TBC. Otherwise, if the two side materials have opposite phonon characteristics, e.g., one is lighter than the middle material and the other is heavier than the middle material, the existence of the third material can decrease the first interface's TBC.

The impact of the second interface on the first interface's TBC depends on the distance between the two interfaces. In Fig. 5 (a), we compare the $G_1$ of Si/Ge and Si/Ge/Si as a function of the distance between the two interfaces. Note that $G_1$ and $G_2$ are the same for Si/Ge/Si since the system is symmetric, and thus, we only discuss $G_1$. We find that $G_1$ in Si/Ge/Si can be 100% higher than $G_1$ in Si/Ge, when the second interface is close to the first interface (distance < 10 nm). As the distance gradually increases, the impact of the second interface gradually decreases and becomes negligible (<10 %) when the distance is greater than 160 nm. In Fig. 5 (b), we compare $G_1$ of Ge/Si and Ge/Si/Ge as a function of the distance between the two interfaces. The impact of the second interface is large when the distance is small and is still notable (~22%) when the distance is 160 nm. This is understandable since the impact of the second interface depends on the MFP of the middle material. If the distance between two interfaces is shorter than the middle material's MFP, the impact of the second interface on the first interface would be significant. For Si/Ge/Si, the characteristic length is Ge's MFP, 102 nm. For Ge/Si/Ge, the characteristic length is Si's MFP, 178 nm.

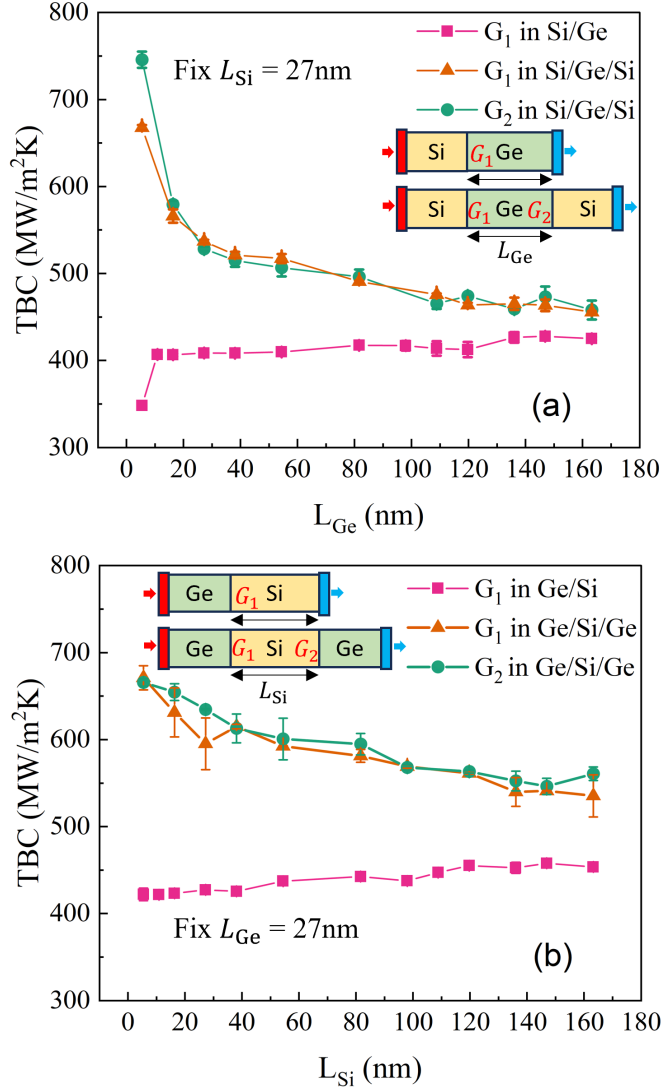

**FIG. 5.** The impact of second interface on first interface's TBC, as a function of distance between two interfaces. (a) $G_1$ of Si/Ge and Si/Ge/Si, as function of Ge thickness. (b) $G_1$ of Ge/Si and Ge/Si/Ge as a function of Si thickness.

## C. Impact of second interface on first material's thermal conductivity

At nanoscales, similar to TBC, the thermal conductivity is not an intrinsic property of a material either, which depends on the material's thickness[58,59,64,72], its neighbor material, and even its second-neighbor material. In this section, we study how a material's thermal conductivity is affected by its neighbor material and its second-neighbor material.

Figure 6 (a-c) show $k_{Si}$ in Si, Si/Ge, and Si/Ge/M, respectively. We find that the first neighbor of Si can significantly increase its thermal conductivity for given thicknesses. For example, $k_{Si}$ in the 27nm-Si/27nm-Ge heterostructure is 200% higher than $k_{Si}$ of a standalone 27nm Si. This is due to the size effect: adding a segment of Ge to Si can mitigate the size effect of Si since some phonons can continue propagation after crossing the interface. We can expect this impact to disappear when the thickness of Si reaches the diffusive limit (>> MFP, 178 nm). The second neighbor's impact is noticeable but smaller compared to the first neighbor's impact. Specifically, $k_{Si}$ in Si/Ge/M is within a 25% difference from $k_{Si}$ in Si/Ge. In summary, one cannot use the thermal conductivity value of standalone semiconductor material to simulate the heat transport inside chips that are made of layers of semiconductors with thicknesses of nano to micrometers. In order words, the thermal conductivities of material A are different in A, A/B, A/C, and A/B/C systems.

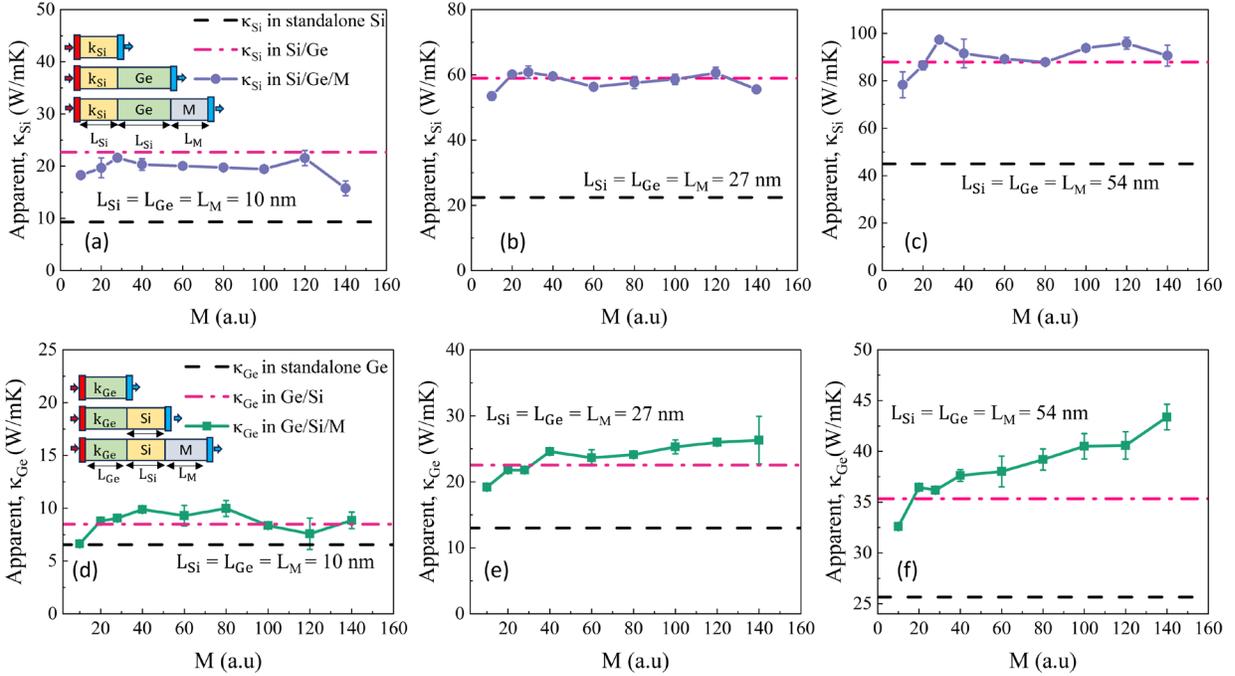

**FIG. 6.** Impact of neighboring and second-neighboring materials on the first material's apparent thermal conductivity. (a-c) $k_{Si}$ in Si, Si/Ge, and Si/Ge/M as a function of mass of M. (d-f) $k_{Ge}$ in Ge, Ge/Si, and Ge/Si/M as a function of mass of M.

Figure 6 (d-f) show $k_{Ge}$ in Ge, Ge/Si, and Ge/Si/M, respectively. Taking $L_{Si} = L_{Ge} = L_M = 54$ nm as an example, the existence of neighboring Si can increase $k_{Ge}$ by 40%, and the existence of the second neighbor, M=140, can increase $k_{Ge}$ by an additional 30%. The impact of the second neighbor is significant. A heavier mass of material excites more acoustic phonons than optical phonons, which can increase the thermal conductivity of the neighboring material[58,69–71]. This might explain (1) why Ge can increase $k_{Si}$ more than Si does to $k_{Ge}$, and (2) why $k_{Ge}$ in Ge/Si/M systems is larger when M is larger. Ge can increase Si's thermal conductivity more than Si can increase Ge's. This is why the thermal conductivity of Ge is higher for heavier M in Ge/Si/M.

Figure 7 shows the dependence of a material's thermal conductivity on the distance from the second neighbor material. On the one hand, taking Si as the studied material, when the Si thickness is kept at 27 nm, and Ge thickness varies from 5 to 160 nm, we find (1) $k_{Si,app}$ in Si/Ge is consistently 120% higher than the standalone Si, (2) $k_{Si,conv}$ in Si/Ge is consistently 200% higher than the standalone Si, (3) $k_{Si,app}$ in Si/Ge/Si is consistently 10% higher than that in Si/Ge, and (4) $k_{Si,conv}$ in Si/Ge/Si is similar to $k_{Si,conv}$ in Si/Ge. On the other hand, taking Ge as the studied material, when Ge thickness is kept at 27 nm, and Si thickness varies from 5 to 160 nm, we find (1) $k_{Ge,app}$ in Ge/Si is consistently 80% higher than the standalone Ge, (2) $k_{Ge,conv}$ in Ge/Si is consistently 10% higher than the standalone Ge, (3) $k_{Ge,app}$ in Ge/Si/Ge is consistently 20% higher than $k_{Ge,app}$ in Ge/Si, and (4) $k_{Ge,conv}$ in Ge/Si/Ge is 40% higher than that in Ge/Si. In summary, the impact of the neighbor and second-neighbor material on the first material's thermal conductivity is non-negligible and nearly distance-independent. Therefore, one can add a thin layer of material onto the original material to tune its thermal conductivity.

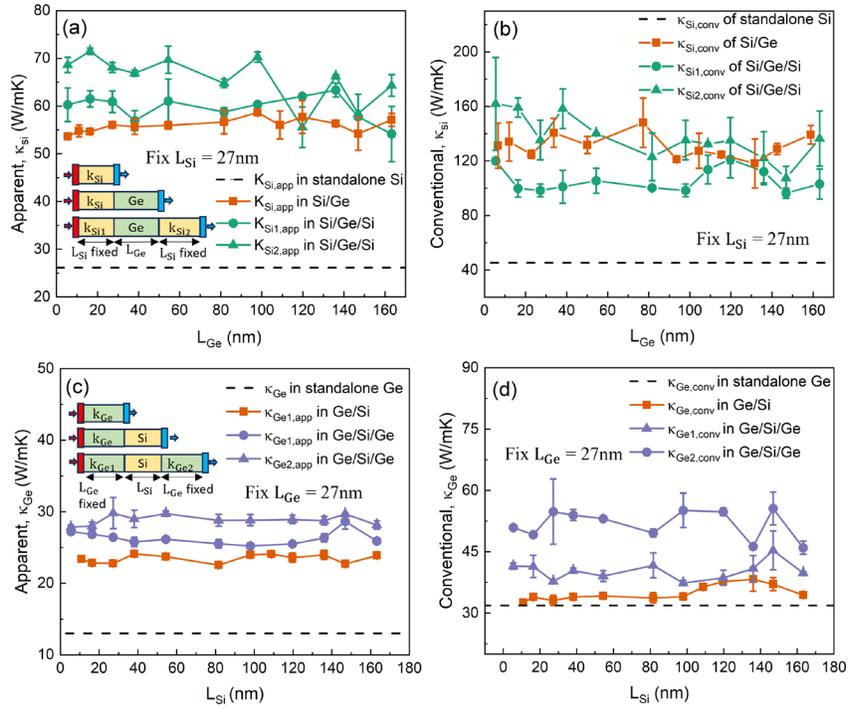

**FIG. 7.** Impact of neighboring and second-neighboring materials on the first material's thermal conductivity, as a function of neighboring material thickness. (a) $k_{Si,app}$ in Si, Si/Ge, and Si/Ge/Si as a function of Ge thickness. (c) $k_{Ge,app}$ in Ge, Ge/Si, and Ge/Si/Ge as a function of Si thickness. (b,d) are the same as (a,c) but for $k_{conv}$.

### D. Impact of the second interface on middle material's thermal conductivity

Figure 8 shows the impacts of neighboring materials on the thermal conductivity of a given material. Figure 8 (a-c) show $k_{Ge}$ in Ge, Si/Ge, and Si/Ge/M as a function of the mass of M. Taking 10-nm Ge as an example, the standalone Ge has a thermal conductivity of 21 W/mK. After attaching to 10-nm Si, $k_{Ge}$ becomes 25 W/mK, a 20% increase. When sandwiched between 10-nm Si and 10-nm M, $k_{Ge}$ is increased up to 50 W/mK, a 1.5-fold increase from standalone Ge, depending on the mass M. Figure 8 (d-f) shows $k_{Si}$ in Si, Ge/Si, and Ge/Si/M as a function of the mass of M. Taking 10-nm Si as an example, the standalone Si has a thermal conductivity of 18 W/mK. After attaching to 10-nm Ge, $k_{Si}$ becomes 53 W/mK, a 3-fold increase. When sandwiched between 10-nm Ge and 10-nm M, $k_{Si}$ is increased up to 175 W/mK, an 8-fold increase, depending on the mass M. In general, no matter if Si or Ge sits in the middle, heavier mass on the two sides can increase the thermal conductivity of the middle material. This is understandable since heavier mass on the two sides filters out high-frequency optical phonons, leaving low-frequency acoustic phonons transporting heat in the middle material. Acoustic phonons have higher thermal conductivity than optical phonons [58,69–71].

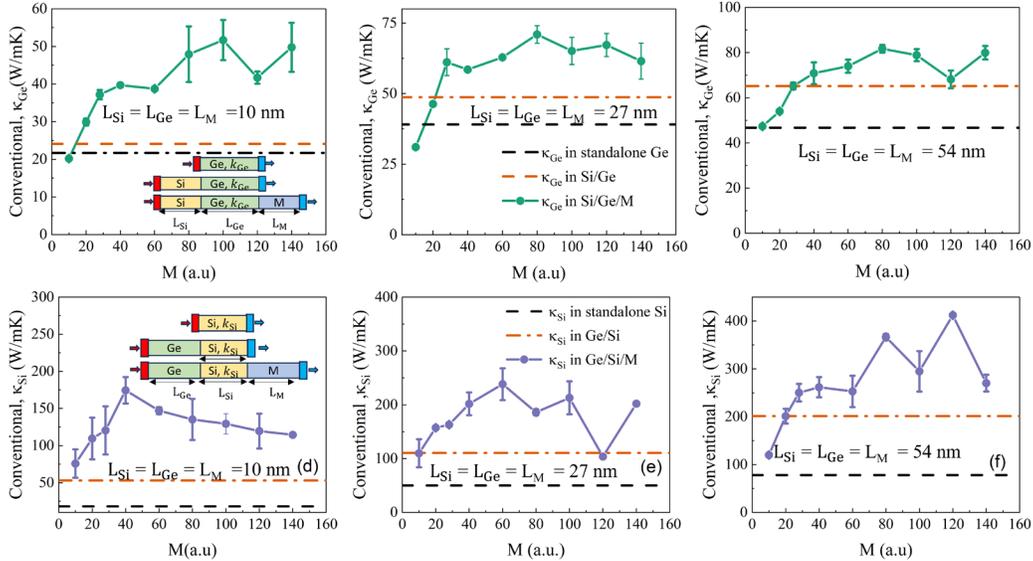

**FIG. 8.** Impact of interfaces on the middle material's thermal conductivity. (a-c) $k_{Ge,conv}$ of Ge, Si/Ge, and Si/Ge/M systems as a function of mass of M. (b-d) $k_{Si,conv}$ of Si, Ge/Si, and Ge/Si/M systems as a function of mass of M.

Figure 9 shows the impacts of the distance between interfaces on the thermal conductivity of the middle material. We take Ge/Si/Ge and Si/Ge/Si as examples. We find that when Ge is sandwiched by two Si, its thermal conductivity is increased but not significantly. This is because the two-side Si excite more high-frequency phonons, which carrier less heat than the low-frequency phonon in Ge. That is, the two side Si are "bad" filters for the middle Ge. In contrast, when Si is sandwiched by two Ge, its thermal conductivity increases by more than two times, even beating its bulk thermal conductivity value. For example, at around 30 nm, the Si's thermal conductivity is increased from its standalone value 50 to 280 W/mK when sandwiched by two Ge slabs. This is because Ge on the two sides filters out high-frequency phonons, leaving more heat-carrying low-frequency phonons in the middle. To gain insight, we extract the mode-dependent heat fluxes $q_\lambda$ of a standalone Si and a Si sandwiched between Ge, where $\lambda$ stands for a phonon mode with certain wavevector and dispersion branch. The methodology is the same as our previous work [68]. We obtain the mode-dependent thermal conductivity $k_\lambda$ using the MD linear region temperature gradient $k_\lambda \equiv q_\lambda / \nabla T_{\text{lin reg}}$. As shown in Fig. 10, after being sandwiched between Ge, the mode-dependent thermal conductivities of Si are significantly higher than those when standalone. The weight of the low-frequency phonons clearly increases. In summary, when a high-frequency material is sandwiched by two low-frequency materials, its thermal conductivity can be boosted to be even higher than the bulk value. This impact will disappear at the diffusive limit, when the length of the middle material is much longer than its MFP.

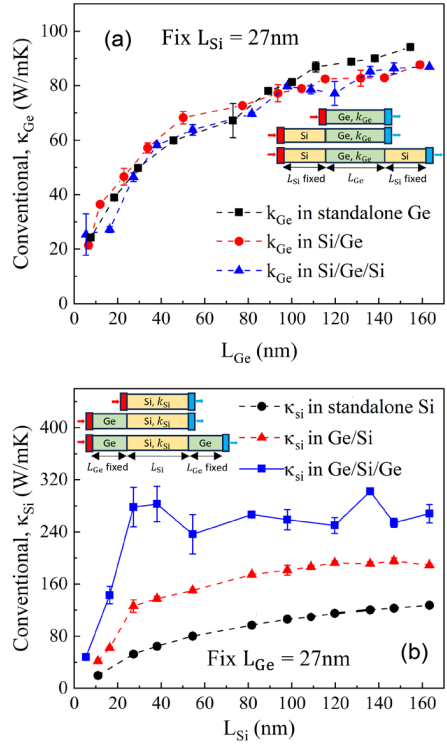

**FIG. 9.** Impact of interfaces on the middle material's thermal conductivity, as a function of the distance between interfaces. (a) $k_{\text{Ge},conv}$ of Ge, Si/Ge, and Si/Ge/Si systems as a function of Ge thickness. (b-d) $k_{\text{Si},conv}$ of Si, Ge/Si, and Ge/Si/Ge systems as a function of Si thickness.

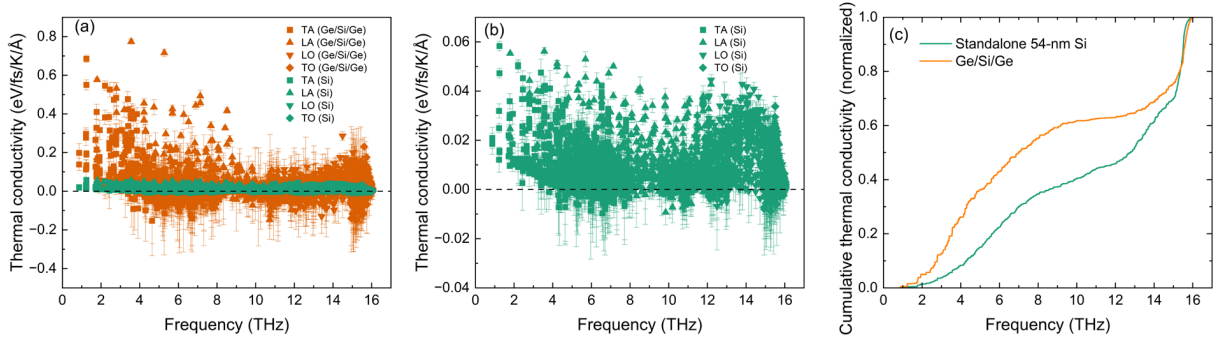

**FIG. 10.** (a) Mode-dependent thermal conductivity of Si in Ge/Si/Ge heterostructure, compared to standalone Si. (b) Mode-dependent thermal conductivity of standalone Si. (c) Normalized cumulative thermal conductivity. In all cases, Si is about 54 nm-thick, and Ge is about 27 nm-thick.

### E. Effect of the second interface on the total thermal resistance of the systems

In the proceeding discussion, we focus on the impact of the third material, M, on the first interface and the first two materials. A natural question is how M impacts the second interface and the total resistance of the system. For example, $G_1$ in Ge/Si/$^{300}$M is larger than that in Si/Ge/$^{300}$M (Sec. III B), but does it mean the total thermal resistance of Ge/Si/$^{300}$M is smaller than that of Si/Ge/$^{300}$M? Similarly, $G_1$ in Si/Ge/$^{6}$M is larger than that in Ge/Si/$^{6}$M (Sec. III B), but does it mean the total thermal resistance of Si/Ge/$^{6}$M is smaller than that of Ge/Si/$^{6}$M? The answer is No. We summarize the thermal resistance components of these systems in Table 1. For example, in Ge/Si/$^{300}$M, $^{300}$M increases the Ge/Si TBC but with the cost of creating an abrupt high-resistance Si/$^{300}$M interface. The abrupt Si/$^{300}$M interface filters out most high-frequency phonons, leaving the transmission at Ge/Si interface high, which increases Ge/Si TBC. Therefore, the increase of an interface's TBC is done with the cost of creating a sharp filter (neighboring interface). As a result, the overall resistance is lower for the heterostructures with smoother transitions. For example, Si/Ge/$^{300}$M has a lower resistance than Ge/Si/$^{300}$M, and Ge/Si/$^{6}$M has a lower resistance than Si/Ge/$^{6}$M.

| Systems | $R_1$ | $R_{i1}$ | $R_2$ | $R_{i2}$ | $R_3$ | $R_{tot}$ |
|---|---|---|---|---|---|---|
| Si/Ge/$^{300}$M | 0.35 | 2.73 | 0.35 | 10.64 | 1.78 | 15.86 |
| Ge/Si/$^{300}$M | 0.58 | 1.44 | 0.16 | 23.62 | 1.42 | 27.22 |
| Ge/Si/$^{6}$M | 0.68 | 4.78 | 0.36 | 3.07 | 0.03 | 8.92 |
| Si/Ge/$^{6}$M | 0.31 | 1.98 | 1.16 | 13.64 | 0.01 | 17.08 |

**Table 1:** Thermal resistance components in Si/Ge/$^{300}$M, Ge/Si/$^{300}$M, Si/Ge/$^{6}$M, and Ge/Si/$^{6}$M systems. Unit of $R$: m$^2$K/GW.

## IV. CONCLUSIONS

In summary, we have studied how much the interplay between two interfaces depends on the distances and interface types, using model systems, i.e., Si/Ge vs. Si/Ge/M and Ge/Si vs Ge/Si/M. The following conclusions are drawn. (1) The TBC of the first interface in double-interface systems is influenced by the second interface, which filters and selectively allows specific phonon modes to pass through. (2) The impact depends on the types of the second interface "Si/M" or "Ge/M", and for most M values, the TBC is enhanced. (3) The interplay will diminish when the distance between interfaces reaches the diffusive limit. The higher MFP of phonons in the middle material leads to a slower diminishing effect of the interplay between interfaces, while lower MFP results in faster attenuation. (4) The presence of the second interface has notable impacts on the thermal conductivity of the first and middle materials. The thermal conductivity of the middle material is influenced by phonon excitation mismatch in the neighboring materials, highlighting the significant role played by interfaces in phonon filtering and excitation mismatch. (5) The total thermal resistance is not just the addition of the "standalone" thermal resistance of constituent materials and interfaces. Caution should be taken when reporting, comparing, or using any measured TBC or thermal conductivity when the system has a nearby interface or material. This study underscores how much the nearby environment can affect the TBC and thermal conductivity, prompting a reevaluation of the thermal resistance models used in thermal characterization and management of nanoscale semiconductor components[81–86]. Note that the interplay revealed in this work is based on lattice-matched Si/Ge model systems with classical potentials. The impact on realistic mismatched and roughened interfaces is expected to be different and warrants future study.

## ACKNOWLEDGMENTS

This work is supported by the National Science Foundation (NSF) (award number: CBET 2337749). We also acknowledge partial support from Oak Ridge Associated Universities (ORAU) Ralph E. Powe Junior Faculty Enhancement Awards. The support and resources from the Center for High Performance Computing at the University of Utah are gratefully acknowledged.

## AUTHOR DECLARATIONS
**Conflict of Interest**
The authors have no conflicts to disclose.

**Author contribution**

## DATA AVAILABILITY

The data that support the findings of this study are available from the corresponding author on reasonable request.

**Author Contributions**

**Khalid Zobaid Adnan**: Investigation (lead); Methodology (supporting); Writing – original draft (equal). **Tianli Feng**: Conceptualization (lead); Funding acquisition (lead); Methodology (lead); Project administration (lead); Supervision (lead); Writing – original draft (equal).